\documentclass[aip,reprint]{revtex4-1}

\usepackage{xcolor}
\definecolor{OliveGreen}{rgb}{0,0.5,0}
\usepackage{graphicx}%
\usepackage{dcolumn}%
\usepackage{bm}%
\usepackage{mathtools}

\draft % marks overfull lines with a black rule on the right

\begin{document}

%\preprint{}

\title{Faraday Rotation Study of Plasma Bubbles in GeV Wakefield Accelerators} %Title of paper



\author{Yen-Yu Chang}
\email[]{Current affiliation:  Helmoltz-Zentrum Dresden-Rossendorf}
%\thanks{}
\affiliation{$^1$Department of Physics, The University of Texas at Austin, Austin, Texas 78712-1081, USA}

\author{Xiantao Cheng}
\affiliation{$^1$Department of Physics, The University of Texas at Austin, Austin, Texas 78712-1081, USA}

\author{Andrea Hannasch}
\affiliation{$^1$Department of Physics, The University of Texas at Austin, Austin, Texas 78712-1081, USA}

\author{Maxwell LaBerge}
\affiliation{$^1$Department of Physics, The University of Texas at Austin, Austin, Texas 78712-1081, USA}

\author{Joseph M. Shaw}
\affiliation{$^1$Department of Physics, The University of Texas at Austin, Austin, Texas 78712-1081, USA}

\author{Kathleen Weichman}
\email[]{Current affiliation:  Laboratory for Laser Energetics}
\affiliation{$^1$Department of Physics, The University of Texas at Austin, Austin, Texas 78712-1081, USA}

\author{James Welch}
\affiliation{$^1$Department of Physics, The University of Texas at Austin, Austin, Texas 78712-1081, USA}

\author{Aaron Bernstein}
\affiliation{$^1$Department of Physics, The University of Texas at Austin, Austin, Texas 78712-1081, USA}

\author{Watson Henderson}
\affiliation{$^1$Department of Physics, The University of Texas at Austin, Austin, Texas 78712-1081, USA}

\author{Rafal Zgadzaj}
\affiliation{$^1$Department of Physics, The University of Texas at Austin, Austin, Texas 78712-1081, USA}

\author{Michael C. Downer}
\email[]{downer@physics.utexas.edu}
\affiliation{$^1$Department of Physics, The University of Texas at Austin, Austin, Texas 78712-1081, USA}


\date{\today}
 
\begin{abstract}

We visualize plasma bubbles driven by 0.67 PW laser pulses in plasma of density $n_e \approx 5\times10^{17}$ ${\rm cm}^{-3}$ by imaging Faraday rotation patterns imprinted on linearly-polarized probe pulses of wavelength $\lambda_{pr} = 1.05\,\mu$m and duration $\tau_{pr} = 2$ ps or $1$ ps that cross the bubble's path at right angles. When the bubble captures and accelerates tens to hundreds of pC of electron charge, we observe two parallel streaks of length $c\tau_{pr}$ straddling the drive pulse propagation axis, separated by $\sim45$ $\mu$m, in which probe polarization rotates by $0.3^\circ$ to more than $5^\circ$ in opposite directions.  Accompanying simulations show that they result from Faraday rotation within portions of dense bubble side walls that are pervaded by the azimuthal magnetic field of accelerating electrons during the probe transit across the bubble. Analysis of the width of the streaks shows that quasi-monoenergetic high-energy electrons and trailing lower energy electrons inside the bubble contribute distinguishable portions of the observed signals, and that relativistic flow of sheath electrons suppresses Faraday rotation from the rear of the bubble. The results demonstrate favorable scaling of Faraday rotation diagnostics to $40\times$ lower plasma density than previously demonstrated.  

\end{abstract}

\pacs{}% insert suggested PACS numbers in braces on next line

\maketitle %\maketitle must follow title, authors, abstract and \pacs


%\label{}
%\subsection{}
%\subsubsection{}

%%%%%%%%%%%%%%%%%%%%%%%%%%%%
%
%		I. Introduction
%
%%%%%%%%%%%%%%%%%%%%%%%%%%%%

\section{Introduction}

Laser-driven plasma accelerators (LPAs), first proposed in 1979,\cite{Taj79} have become tabletop sources of quasi-monoenergetic GeV electron bunches \cite{Gon19} and ultrafast secondary X-ray pulses \cite{Cor13}, for applications in biology, medicine, and materials science.\cite{Hoo13} The key to the compact size of LPAs is the unprecedented accelerating field ($\sim$GV/cm), contained in a light-speed, microscopic (tens of $\mu$m) plasma structure, which contrasts with $<$ MV/cm fields to which metal cavities of conventional radio-frequency accelerators are limited. The highest-performing LPAs operate in a strongly nonlinear "bubble" regime,\cite{Lot04,Puk02,Bar04,Kos04,Lu06} in which the driving laser pulse is shorter than a plasma period $\omega_p^{-1}$, where $\omega_p$ is the plasma frequency, and intense enough to blow out electrons completely from its immediate wake, forming a near-spherical ion cavity lined by a thin, dense wall (or sheath) of electrons. 

Optical diagnostics that visualize the transient, evolving structure of LPAs, and link it to accelerated electron beam (\textit{e}-beam) properties and to theory and simulation, have played an important role in advancing LPA science.\cite{Dow18}  Previous diagnostic experiments have visualized laser-driven plasma bubbles in plasmas of near-atmospheric electron density ($n_e\sim 10^{19}$ cm$^{-3}$), for which dephasing of accelerating electrons from the plasma bubble and erosion of the drive pulse limit acceleration length to a few mm, and final electron energy to $\sim 100$ MeV.\cite{Lu07}  These experiments utilized the ability of a plasma bubble in dense plasma to refract and re-shape a co-propagating near-ultraviolet (UV) probe,\cite{Don10} to phase-modulate an oblique-angle near-UV probe,\cite{Li14} or to Faraday-rotate\cite{Kalu10,Buc11} or refract\cite{Sav15} a transverse near-infrared (IR) probe.  

Forefront LPAs that accelerate electrons to multi-GeV energy,\cite{Gon19,Wan13,Lee14} on the other hand, require much lower density plasma ($n_e \sim$ few\,$10^{17}$ cm$^{-3}$) in order to extend dephasing and pump depletion lengths to multiple centimeters.\cite{Lu07}  To form plasma bubbles, they also require laser drivers of peak power $P \agt 10P_{cr}$,\cite{Kalm10} where $P_{cr} = 17(n_{cr}/n_e)$ GW is the critical power for relativistic self-focusing and $n_{cr} \approx 10^{21}/[\lambda(\mu{\rm m})]^2$ cm$^{-3}$ is the critical plasma density for a driver of wavelength $\lambda$. This enables the drive pulse to self-focus smoothly to field strengths $3 \alt a_0 \alt 6$ sufficient to blow out a steady-state bubble.  For a Gaussian drive pulse focused in a ``matched'' geometry, in which drive pulse and bubble propagate without oscillating transversely, the bubble radius is\cite{Lu07} \label{Rb}
\begin{equation}
R_b = (\lambda_p/\pi)\sqrt{a_0} \approx \sqrt{\frac{a_0}{n_e ({\rm cm}^{-3})}} \times 10^{10} \mu {\rm m}.  
\end{equation}
Here, $a_0 = eE_l/mc\omega$ is a dimensionless laser strength parameter related to peak intensity $I_0$ by $a_0 = 0.85\,\lambda(\mu{\rm m})\, I^{1/2}_0(10^{18}\,{\rm W\,cm}^{-2})$, $\lambda_p = 2\pi c/\omega_p$ is the plasma wavelength, and $E_l$ and $\omega$ are laser electric field and frequency, respectively.  Thus e.g. at $n_e = 5 \times 10^{17}$ cm$^{-3}$, a drive pulse of $P \agt 0.35$ PW at $\lambda = 1\,\mu$m is needed to blow out a plasma bubble, which from Eq.\,(1) has radius $25 < R_b < 35 \mu$m upon self-focusing to $3 < a_0 < 6$.  In contrast, at $n_e = 10^{19}$ cm$^{-3}$, a $\lambda = 1\,\mu$m drive pulse of $P \agt 18$ TW suffices, and blows out a bubble of radius $6 < R_b < 8 \mu$m upon equivalent self-focusing. 

Large bubbles at low $n_e$ offer a potential diagnostic opportunity to resolve finer bubble sub-structure than was possible in high-$n_e$ experiments,\cite{Don10,Li14,Kalu10,Buc11,Sav15} if they could be observed at equivalent near-IR/UV probe wavelengths $\lambda_{pr}$. However, the low refractive index contrast $\eta - 1 \sim 10^{-4}$ between the bubble's interior ($\eta = 1$) and surrounding $n_e \approx 5 \times 10^{17}$ cm$^{-3}$ plasma ($\eta \approx  1 - n_e/2n^{(pr)}_{cr}$) at these $\lambda_{pr}$ has prevented optical visualization.  Here, $n^{(pr)}_{cr}$ denotes the critical density at $\lambda_{pr}$.  One can recover optical contrast by using mid-IR probe pulses that maintain equivalent $n_e/n^{(pr)}_{cr}$, but the advantage of higher feature resolution is then lost.\cite{Dow18}  Alternatively, Zhang et al.\cite{Zha17} used fs \textit{e}-bunches from a separate ``diagnostic'' LPA to probe internal structure of laser-induced wakes in plasma of $n_e$ as low as $3\times 10^{17}$ cm$^{-3}$. However, the detected plasma wave was in the linear, rather than bubble, regime.  Moreover, adding a source of synchronized fs \textit{e}-bunches complicates the diagnostic setup significantly.

%%%%%%%%%%%%%%%%%%%%%%%
%
%			FIG 1
%
%%%%%%%%%%%%%%%%%%%%%%%

\begin{figure}
\includegraphics[width=0.5\linewidth, trim={0cm 0cm  0cm 0cm},clip]{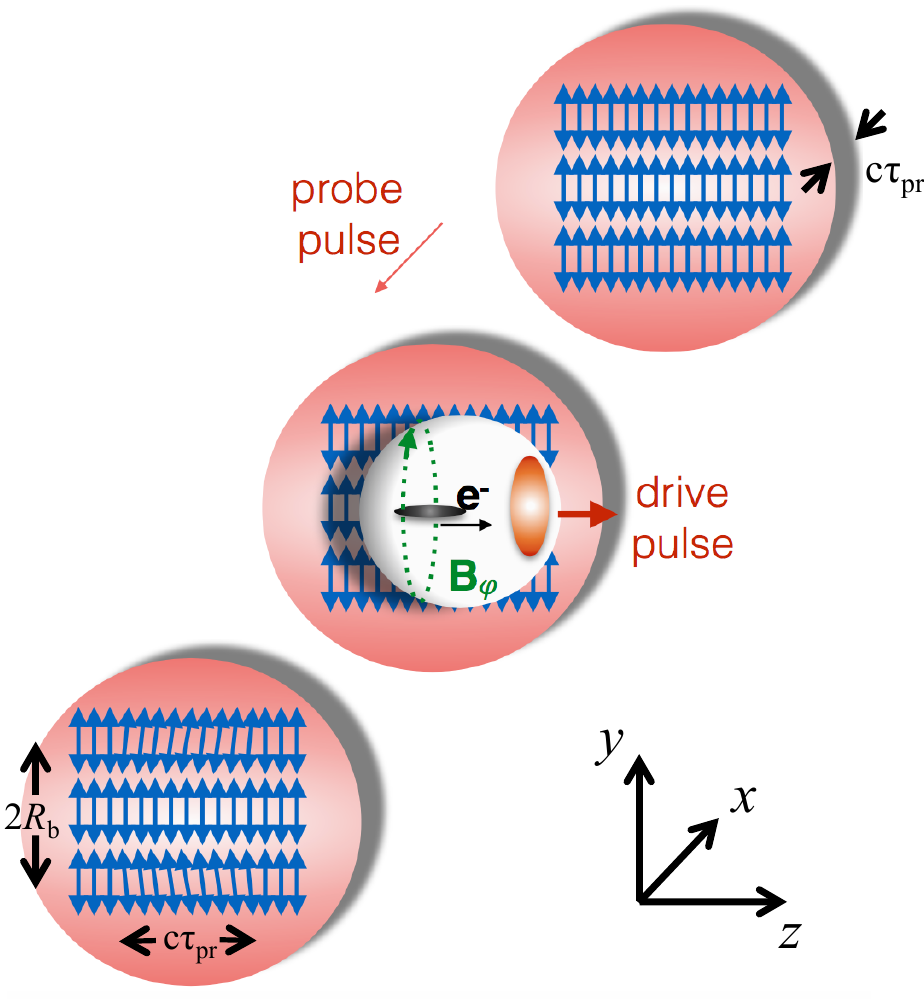}
\caption{Conceptual depiction of local Faraday rotation of polarization of an $x$-propagating probe pulse in those portions of the dense upper and lower walls of a $z$-propagating plasma bubble that are magnetized along $\mp\mathbf{x}$ by the field of captured, accelerating electrons.}
\end{figure}

Here, we optically visualize laser-driven plasma bubbles of radius $R_b \approx 25 \mu$m generated in $n_e \approx 5\times 10^{17}$ cm$^{-3}$ plasma
%  ---  $20$ times lower than in previous work\cite{Kalu10,Buc11,Sav15} ---  
 by imaging Faraday rotation patterns within transversely-propagating probe pulses of \textit{near-IR} wavelength ($\lambda_{pr} = 1.05 \mu$m)
%  --- similar to $\lambda_{pr}$ in previous work --- 
  that are simply split from the drive pulse. %Faraday rotation signatures are observed only in conjunction with self-injection, trapping and acceleration of $\sim 100$ pC of electrons to GeV energy.\cite{Wan13}  
The probe polarization rotation angle is\cite{Kalu10}

\begin{equation}
\phi_{rot}=\frac{e}{2m_ecn^{(pr)}_{cr}}\int_\ell n_e\mathbf{B}_{\varphi}\cdot d\mathbf{s},
\label{rotation_rest}
\end{equation}
where $\mathbf{B}_\varphi$ is the azimuthal magnetic field created by accelerating electrons (with additional contributions from plasma currents), and path element $d\mathbf{s}$ is integrated along the path $\ell$ of each probe ray through the bubble. Probe polarization thus rotates preferentially in regions where $\mathbf{B}_\varphi$ is either parallel or anti-parallel to the probe wavevector $\mathbf{k}_{pr}$, and where $n_eB_\varphi$ is maximum, i.e.\,\,in dense plasma bubble walls, just above and below the accelerating \textit{e}-bunch (see Fig.\,1).   Eq.\,\eqref{rotation_rest} contains the same scaling factor $n_e/n^{(pr)}_{cr}$ --- here $\sim 40 \times$ smaller than in high-$n_e$ Faraday rotation experiments\cite{Kalu10,Buc11} --- that has previously inhibited visualization of low-$n_e$ wakes with near-IR probes.  Here, however, we compensate this deficit with larger $B_\varphi \propto N_e\gamma_e/\tau_bR_b^2$, where $N_e$, $\gamma_e$ and $\tau_b$ are the number, Lorentz factor and duration, respectively, of the accelerated electron bunch, and with longer integration path $L_{int} \sim R_b$, yielding $B_\varphi L_{int} \propto N_e\gamma_e/\tau_bR_b$.  Since here $N_e$, $\gamma_e$ and $R_b$ are larger by factors of $5-500$, $30$ and $\sqrt{40}$, respectively, than in high-$n_e$ Faraday rotation experiments, $B_\varphi L_{int}$ is $\sim25$ to $2500\times$ larger for equivalent $\tau_b$, compensating the smaller $n_e/n^{(pr)}_{cr}$.  As a result, we observe $\phi_{rot}$ ranging from $0.3^\circ$ to $6^\circ$, i.e. up to $\sim20\times$ larger than observed in MeV-class LPAs.  Such large rotations open the possibility of high-resolution probing of bubble morphology and dynamics. 

Section II describes the experimental setup. Section III presents experimental results. Section IV models the experimental results using particle-in-cell (PIC) simulations of plasma bubble structure and finite-difference time-domain (FDTD) simulations of Faraday rotation.  We state our conclusions in Section V. 

%%%%%%%%%%%%%%%%%%%%%%%%%%%
%
%		II. Experimental Procedure
%
%%%%%%%%%%%%%%%%%%%%%%%%%%%

\section{Experimental Procedure}

Fig.\,2(a) shows the experimental setup schematically. A spherical mirror focused a drive pulse with $P = 0.67$ PW ($100$ J, $150$ fs FWHM) and $\lambda = 1.057$ $\mu$m from the Texas Petawatt Laser\cite{Gau16} at $f/40$ into the entrance aperture of a $7$-cm-long rectangular-prism-shaped gas cell filled with 5 Torr helium (He) of $99.99\%$ purity. The drive pulse ionized the gas, producing plasma of density $n_{e} \approx 5 \times10^{17}$ cm$^{-3}$, then self-focused and generated a nonlinear plasma wave, which captured and accelerated electrons to GeV energy.  A 1T magnetic field deflected electrons in a plane perpendicular to the drive laser polarization onto an imaging plate. Electron spectra contained a quasi-monoenergetic peak with $\sim10$ to $>1000$ pC charge at energies ranging from 0.6 to 1 GeV, depending on laser-plasma conditions, and a low-energy tail [Fig.\,2(a), top right].  See Wang \textit{et al.}\cite{Wan13} for further details.  

%%%%%%%%%%%%%%%%%%%%%%%%%%%
%
%			FIG 2
%
%%%%%%%%%%%%%%%%%%%%%%%%%%%

\begin{figure}
\includegraphics[width=0.85\linewidth, trim={0cm 0cm  0cm 0cm},clip]{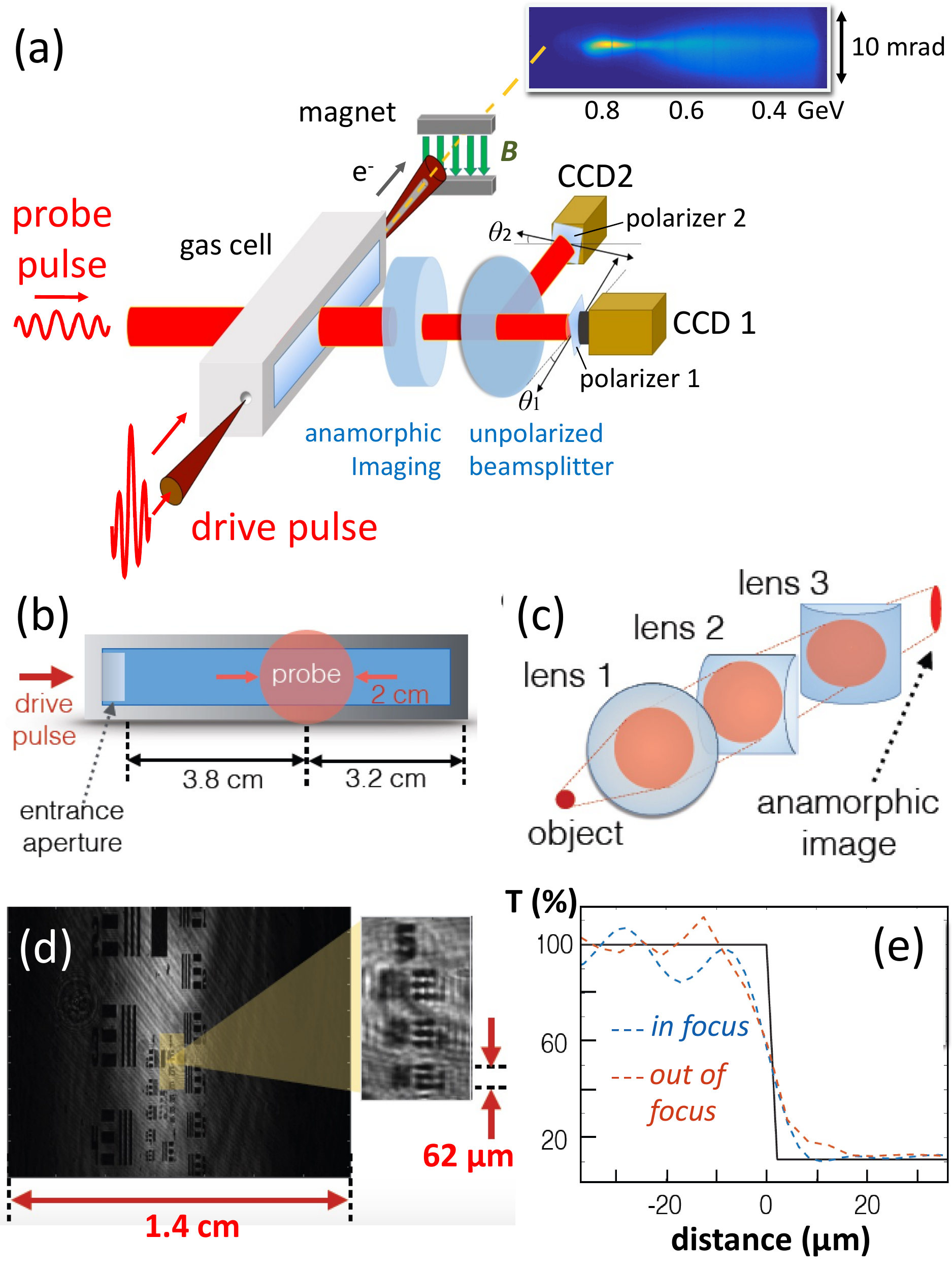}
\caption{ \label{setup} Schematic of GeV LPA and Faraday rotation diagnostics. (a) Overall layout. (b) Gas cell, side view. Drive and probe pulses are both vertically polarized.  (c) Anamorphic imaging optics.  Lens 1: spherical achromatic doublet, $f_1=15$\,cm, 2-inch diameter; Lenses 2, 3: cylindrical achromatic doublets, $f_{2,3}=25$\,cm, $3$\,cm. (d) Anamorphic image of Air Force test resolution chart with $1.4$\,cm (0.25\,cm) horizontal (vertical) field of view. (e)  Detail of horizontally-averaged $\%$ probe transmission vs.\,vertical distance from horizontal edge (black solid) of test chart element when in focus (blue-dashed), and $300$ $\mu$m out of focus (dashed red), demonstrating $12.5\,\mu$m vertical resolution, $300$ $\mu$m depth of field.}
\end{figure}

For Faraday rotation studies, we split a probe pulse of $\lambda_{pr} = 1.057 \mu$m from a preamplifier stage of the laser system, compressed it to duration $\tau_{pr} \approx 2$\,ps or $1$\,ps and telescoped it to radius $w^{(pr)}_0 \approx 1$\,cm. The collimated probe, co-polarized with the pump, entered the gas cell through a rectangular side window, with its center at distance $z = 3.8$\,cm from the cell entrance [Fig.\,\ref{setup}(b)].  The probe exited into a detection system through a matching window on the opposite side [Fig.\,\ref{setup}(a)].  This crossing point avoided a region of strong pump side-scatter in the first 2-3 cm of the cell,\cite{Wan13} while probing a region in which, according to simulations,\cite{Kalm10} the drive pulse had finished self-focusing and drove a steady-state bubble.  We diagnosed pump-probe spatiotemporal overlap during test shots with reduced-energy ($\sim$$1$\,J), rep-rated pump pulses and an air-filled cell, by observing probe refraction from the pump-ionized plasma column.  No such probe shadowgraphs were observed, however, when the cell was filled with only 5 Torr He, regardless of pump energy, highlighting the difficulty of transverse shadowgraph diagnostics at this $n_e$ and $\lambda_{pr}$.  We chose $c\tau_{pr}, w^{(pr)}_0 >> R_b$ to ensure pump-probe overlap during infrequent (once per hour) low-$n_e$ full system shots.  Consequently Faraday rotation signals are streaks along $z$ of length $c\tau_{pr} \approx 25R_b \approx 0.6$\,mm (for $\tau_{pr} = 2$\,ps) or $12.5R_b \approx 0.3$\,mm (for $\tau_{pr} = 1$\,ps), determined by probe transit time across the bubble's path. 

Transverse imaging of GeV LPAs requires a multi-cm horizontal field of view, to capture a significant portion of the bubble's propagation distance,\cite{Lu07,Kalm10} together with high vertical resolution, to see bubble features.  To reconcile these competing requirements, we imaged the bubble to detectors anamorphically, demagnifying the horizontal, while magnifying the vertical, dimension.  Fig.\,\ref{setup}(c) shows the anamorphic imaging system, which consisted of a spherical achromatic collecting lens (Lens 1) followed by two orthogonal achromatic cylindrical lenses (Lenses 2, 3).  Images of an Air Force resolution test chart [Fig.\,\ref{setup}(d)] showed that the system provided 1.4 cm ($2.5$\,mm) field of view with $\sim 50$ $\mu m$ ($\sim 12.5$ $\mu$m) resolution horizontallly (vertically), and $300$ $\mu$m depth of field [Fig.\,\ref{setup}(e)].  Since the pump propagation axis in the cell fluctuated laterally $<\,100\,\mu$m RMS from shot to shot, a single pre-run focus adjustment with the test chart sufficed to guarantee in-focus images throughout a run. 

To detect probe polarization rotation with high signal-to-noise ratio, we used a differential detection procedure developed by previous researchers.\cite{Kalu10,Buc11}  A non-polarizing beamsplitter distributed replicas of the wake-modulated probe beam after the anamorphic imaging system to two charge-coupled device cameras [CCD1,2; Fig.\,\ref{setup}(a)] thru polarizers 1,2 that were rotated by small bias angles $\theta_1 = -\theta_2 \approx 2^\circ$ away from extinction in opposite directions. Consequently, regions where probe polarization rotates brightened on one camera and dimmed on the other, as long as the local rotation angle $\phi_{rot}(y,z) < \theta_i$.  From Malus' law, intensity $I_i(y,z)$ on CCD$i$ ($i = 1,2$) is 
\begin{equation}
I_i(y,z) = T_i\,I_0(y,z)[1-\beta_i\cos^2(\phi_{rot}(y,z)-\theta_i)] ,
\end{equation}
where $T_i$ is the transmission/reflection ratio of the beam splitter, $I_0(y,z)$ is the intensity profile of the probe beam incident on the beamsplitter, and $\beta_i$ is the extinction ratio of polarizer $i$ $(1- \beta_1= 6.1 \times 10^{-3}$; $1- \beta_2= 3.1 \times 10^{-3})$. Since for most shots $\phi_{rot}(y,z)$ was clearly less than $\theta_i$, we extracted $\phi_{rot}(y,z)$ simply by dividing $I_1/I_2$ after the two images were mutually aligned using a calibrated reference point.  For occasional high-$N_e$ shots for which $\phi_{rot}(y,z)$ exceeded $\theta_i$, we extract $\phi_{rot}(y,z)$ using an iterative procedure described in the Supplementary Material.  

%%%%%%%%%%%%%%%%%%%%%%%%%%%%%%
%
%		III. Experimental Results
%
%%%%%%%%%%%%%%%%%%%%%%%%%%%%%%

\section{Experimental Results}

\begin{figure}
\centering
\includegraphics[width=\linewidth, trim={0cm 0cm  0cm 0cm},clip]{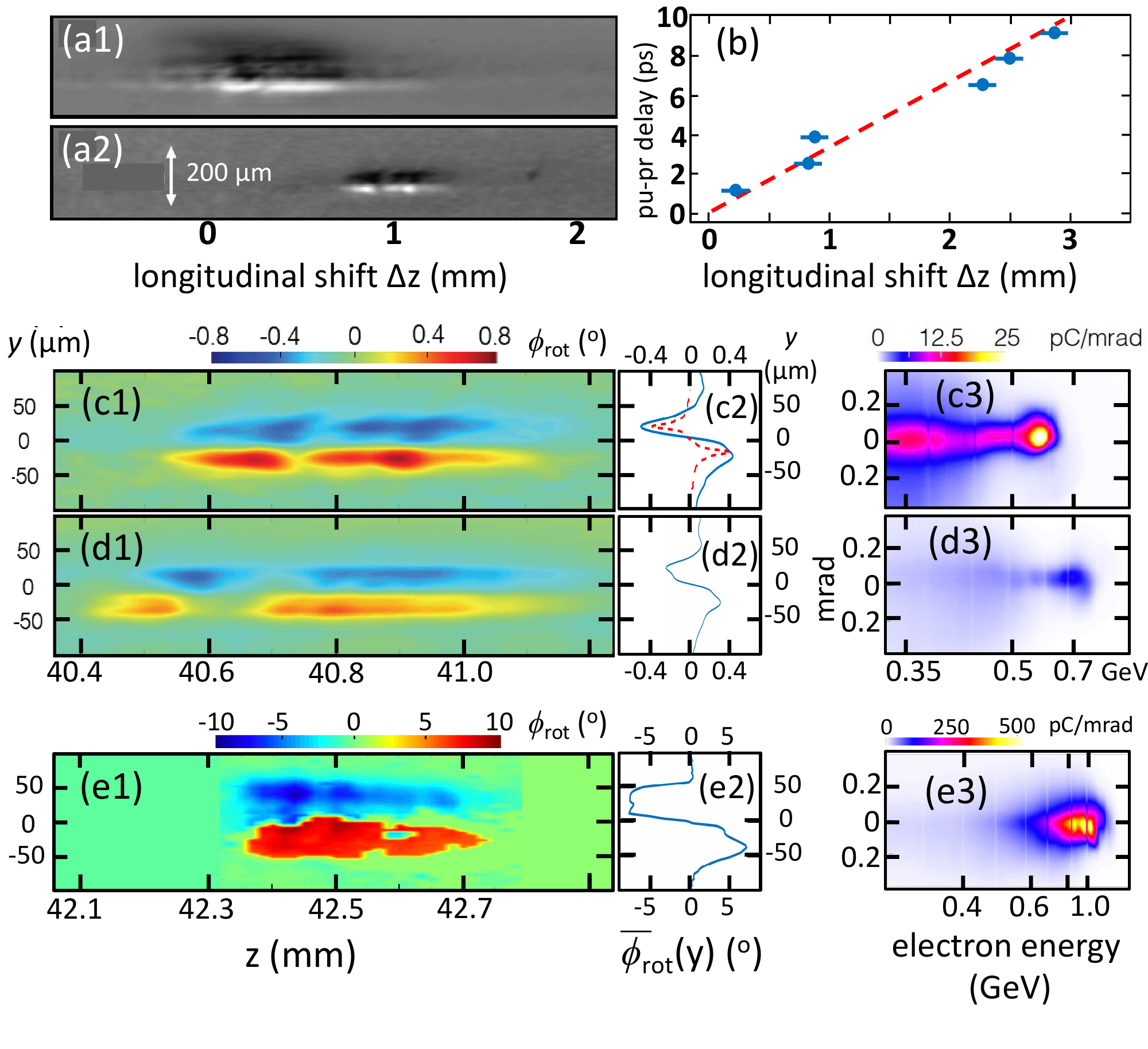}
\caption{Faraday rotation streaks. (a) Differential Faraday rotation images for pump-probe delay $T$ (a1), and $T + 1.2$\,ps (a2), showing shift $\Delta z = (1.2\,{\rm ps})c$ of image along $z$.  (b) Plot of variation $\Delta t$ in pump-probe delay vs.\,streak centroid position $z$.  Red-dashed line: slope $1/c$.  (c1)-(e1): Differential Faraday rotation images, showing $\phi_{rot}(y,z)$ (color scale); $y = 0$ is the drive pulse propagation axis, $z = 0$ is the gas cell entrance. Probe duration (delay) was 2 ps ($T$) for (c2) and (d2), 1 ps ($T+5.7$\,ps) for (e2). (c2)-(e2): plots of $\bar{\phi}_{rot}$ vs.\,$y$, averaged along $z$ (blue solid curves). %The widths (FWHM) of the peaks indicated by grey arrows are (c2) $35$ $\mu$m, .... 
Red dashed curve in (c2): FDTD simulation.  (c3)-(e3):  electron spectra corresponding to (c1)-(e1). Total charge in pC above $300$ MeV:  (c) $46$; (d) $13$; (e) $1390$. 
%(c1), (c2) The normalized signal images that were used to generate (b1). (d) The cartoon of the Faraday effect induced by the plasma bubble which occurs preferentially around the region that overlaps with the plasma bubble wall. 
%(f) The horizontal line-out of (b1) averaged from $y=-40$ to $0$ $\mu$m.
}
\label{data_1}
\end{figure}

Fig.\,\ref{data_1}(a) shows differential Faraday rotation data for two consecutive shots differing by $\Delta t = 1.2$\,ps in pump-probe delay.  The length of each streak is consistent with probe pulse length $c\tau_{pr} \approx 0.6$\,mm, and the lateral shift $\Delta z$ between shots is consistent with $c\Delta t \approx 0.36$\,mm.  Fig.\,\ref{data_1}(b) plots $\Delta t$ vs.\,$\Delta z$ for six shots for which $\Delta t$ varied by 8\,ps.  The fitted slope is $1/c$.  No signal was observed with the probe beam blocked.  These checks demonstrated that observed streaks indeed arose from pump-probe overlap.  

Figs.\,\ref{data_1}(c1),(d1) show differential Faraday rotation images, and Figs.\,\ref{data_1}(c3),(d3) corresponding electron spectra, for two shots using a 2 ps probe at the same delay $T$. Figs.\,\ref{data_1}(e1) and (e3) show the same for a 1 ps probe at delay $T+5.7$\,ps.  In the rotation images, the drive pulse propagates left to right along the $+z$-axis (i.e.\,$y = 0$), and the probe pulse propagates out of the page in the $-\hat{x}$ direction.  Each image consists of two parallel streaks of length $0.6$\,mm [panels (c1),(d1)]  or 0.3\,mm [panel (e1)] straddling the laser propagation axis.  The upper (blue) streak, centered at $y \approx +22\,\mu$m, corresponds to clockwise rotation for an observer facing into the oncoming pulse (i.e.\,$\phi_{rot} < 0)$, as expected when $\mathbf{B}$ points in the $+\hat{x}$ direction; the lower (red) one, centered at $y \approx -22\,\mu$m, corresponds to counter-clockwise rotation ($\phi_{rot} >0$), as expected when $\mathbf{B}$ points in the $-\hat{x}$ direction. These field directions match those of the azimuthal field of a right-propagating \textit{e}-bunch (see Fig.\,1).  The average transverse distance between the centroids of the upper and lower streak is $\Delta y \approx 45$ $\mu$m.  
%, which matches the plasma bubble diameter $2R_b \simeq 56.4$ $\mu$m [see Eq.\,(2)] expected for $n_e = 5 \times 10^{17}$ cm$^{-3}$ and a drive pulse propagating with steady-state field strength $a_0\simeq 3$.  
Figs.\,\ref{data_1}(c2)-(e2) (solid blue) show $z$-averaged vertical line-outs $\bar{\phi}_{rot}(y)$ for the data in Figs.\,\ref{data_1}(c1)-(e1), respectively, together with a simulation result [dashed red, panel (c2)] discussed below.  

%%%%%%%%%%%%%%%%%%%%%%%%%%%%%%
%
%		FIG 3
%
%%%%%%%%%%%%%%%%%%%%%%%%%%%%%%

From the horizontal scales of Fig.\,3(c2)-(e2), the values of $\bar{\phi}_{rot}(y_{max})$ at its peaks $\pm y_{max}$ ranged from $\pm 0.3^\circ$ [Fig.\,3(d2)], for a shot in which the electron spectrum contained only 13 pC above 300 MeV, to $\pm 6^\circ$ [Fig.\,3(e2)], for a shot that accelerated $1.4$\,nC above 300 MeV. Along $z$, the maximum rotation $\phi_{rot}(y_{max},z)$ varied from $\sim\bar{\phi}_{rot}(y_{max})/2$ to $\sim2\bar{\phi}_{rot}(y_{max})$. It is not clear whether these $z$-variations arise from physical variations in bubble characteristics or from uncharacterized spatiotemporal structure in the probe pulse.  Nevertheless, the locations $y_{max}$ of the $\phi_{rot}$ maxima remain nearly constant with $z$ for each shot, indicating that $R_b$ remains constant over the propagation section captured in each shot.  

%Fig.\,3f3 shows the electron spectrum for a shot in which accelerated electrons emerged in two bunches separated by $\sim1$ mrad in a direction perpendicular to the spectrometer's energy dispersion plane. We observe such bifurcation occasionally when the drive laser focuses to a double-peaked intensity envelope, causing it to split into two filaments that drive parallel self-injected bubbles. The corresponding Faraday rotation signal (Fig.\,3f1) shows broader structure in the bifurcation direction ($y$) than single-bubble shots, indicative of the ``double-bubble'' structure of the interaction region.

%%%%%%%%%%%%%%%%%%%%%%%%%%
%
%		IV. Simulations and Discussion
%
%%%%%%%%%%%%%%%%%%%%%%%%%%

 \section {Simulations and Discussion}
 
 %%%%%%%%%%%%%%%%%%%%%%%%%%
%
%		Fig. 4
%
%%%%%%%%%%%%%%%%%%%%%%%%%%
\begin{figure}
\centering
\includegraphics[width=\linewidth, trim={0cm 0cm  0cm 0cm},clip]{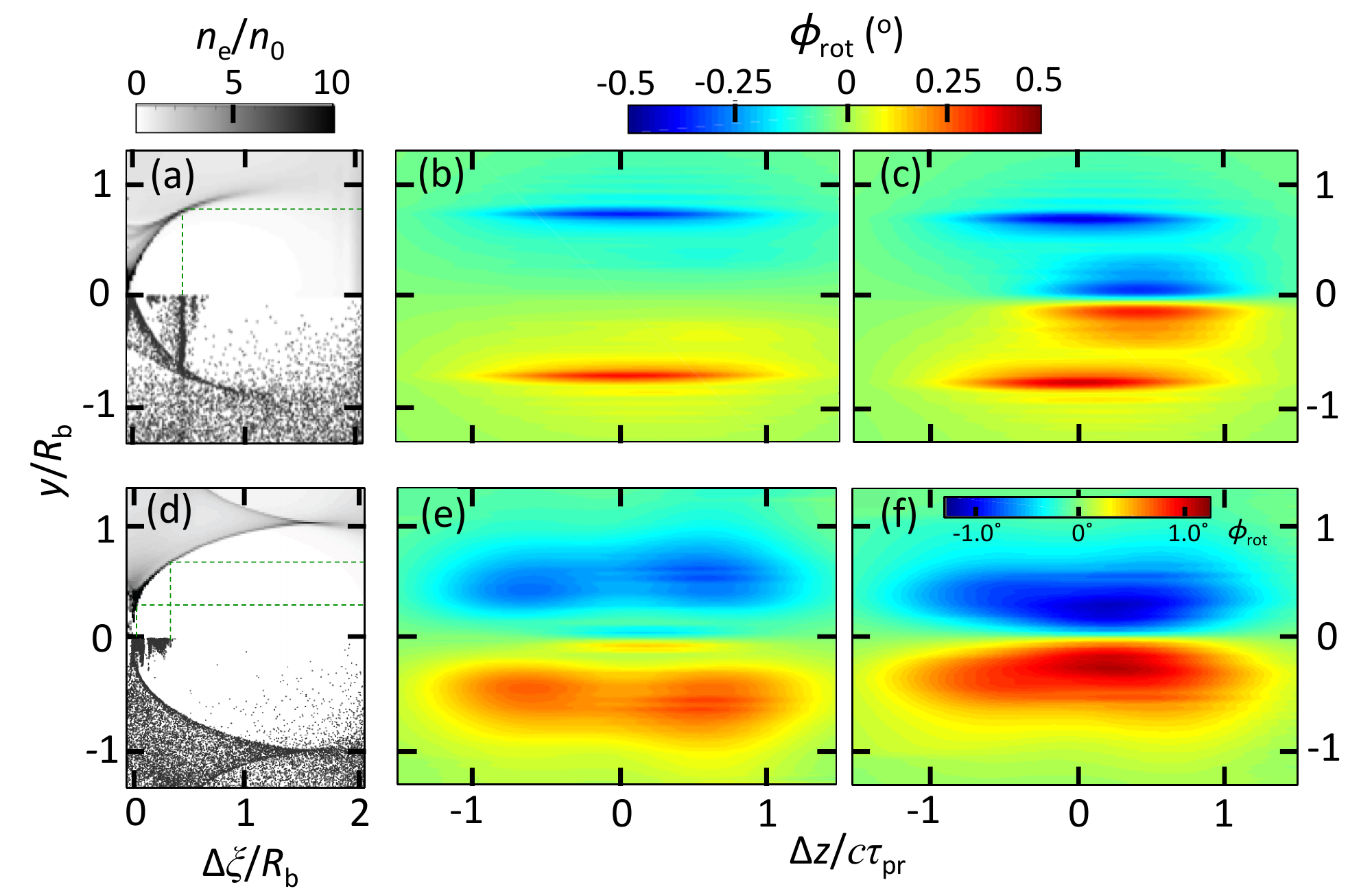}
\caption{\label{2D+1_FDTD} Simulation results. (a) Plasma density profile (upper half) ($n_0=5\times 10^{17}$ $cm^{-3}$) and test particle distribution (lower half) of plasma bubble at $z \approx 40$\,mm from PIC simulation using code WAKE. (b) Calculated Faraday rotation streaks on probe pulse after propagating through the bubble in (a). (c) As in (b), but neglecting relativistic electron flow in bubble walls.  (d) As in (a), but for bubble at smaller $z$, shortly after injection, with low-energy electrons trapped near rear of bubble.  (e) As in (b), after probe propagation through bubble in (d).  (f) As in (e), but neglecting relativistic electron flow in bubble walls.  Color scale is different for (f).}
\end{figure}
 
 In order to analyze observed Faraday rotation streaks quantitatively, we simulated the interaction of a transverse probe pulse with the bubble structure.  
For the latter, we used representative snapshots of the wake's density $n_e(r,\xi, z)$, magnetic field $B_\varphi(r,\xi,z)$ and electron velocity flow field $\bm{\beta}_f(r,\xi,z)$ profiles in the probed $z$-range $3 \alt z \alt 5$\,cm from previously-reported 3D particle-in-cell (PIC) simulations of the Texas Petawatt LPA.\cite{Kalm10,Wan13} The simulations were based on a hybrid approach.  The fully relativistic code WAKE\cite{Mor97} in axisymmetric geometry simulated self-consistent propagation of the drive laser pulse and plasma wake.  The quasi-cylindrical 3D-PIC code CALDER-Circ\cite{Lif09} provided fully dynamic, self-consistent simulations of electron self-injection. A relativistic, fully 3D dynamic, non-averaged test-particle tracking module within WAKE modeled electron acceleration after injection. Further details of the simulation procedure are provided elsewhere.\cite{Kalm10}   In the arguments of $n_e$, $B_\varphi$ and $\bm{\beta}_f$, $r$ denotes transverse distance from the bubble's propagation axis, and $\xi = z - v_bt$ the longitudinal distance from a reference point at the back of the bubble moving at velocity $v_b$. $\bm{\beta}_f$ denotes local flow velocity $\bm{v}_f$ normalized to $c$.  These simulations used experimental plasma and laser parameters reported here, albeit with idealized Gaussian longitudinal and transverse drive pulse profiles.  As in the experiment, drive pulse focus was non-matched, resulting in $\sim\,20\%$ variations in $R_b$ over the probed $z$-range.  Since these variations depend on details of the actual pulse profile, the simulation cannot predict $R_b$ at a specific probe location \textit{ab initio}. The goal is rather to determine the separation and structure of the Faraday rotation streaks for a given simulated $R_b$.  By comparing these to measurements, we then infer the actual $R_b$, and other bubble characteristics, at that location from the data.  The top half of Fig.\,\ref{2D+1_FDTD}(a) shows a representative simulated bubble structure within the probed $z$-range.  Since the bubble is not perfectly spherical, we labeled its maximum transverse extent $R_b$.  The simulation showed $20 \alt R_b \alt 25\,\mu$m in the probed $z$-range.  The bottom half of Fig.\,\ref{2D+1_FDTD}(a) shows test particles, representing injected and accelerating electrons, distributed within the bubble.  The concentration of them at $\Delta\xi/R_b \approx 0.5$ contains $\sim\,40$\,pC above 300\,MeV, and represents the main part of the accelerating $e$-bunch.  Since this is close to the charge above above 300\,MeV observed for the shot that yielded the Fig.\,\ref{data_1}(c) data, we used the structure in Fig.\,\ref{2D+1_FDTD}(a) to model the Faraday streaks in Fig.\,\ref{data_1}(c1).  The smaller (larger) accelerated charge for the Fig.\,\ref{data_1}(d) [Fig.\,\ref{data_1}(e)] data indicates that self-injection dynamics differed substantially from those in the simulation. 

\begin{figure}
\centering
\includegraphics[width=0.75\linewidth,trim={2cm 4cm 2cm 4cm},clip]{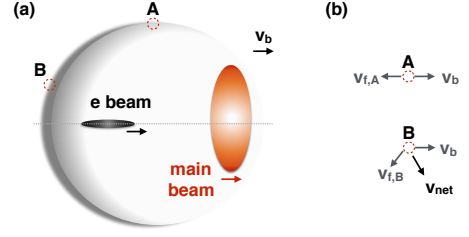}
\caption{\label{bubble_flow} Diagram of plasma bubble and the plasma flow. (a) plasma bubble induced by a laser pulse. (b) qualitative explanation of the plasma net flow at different positions of the bubble wall in the lab frame.  }
\end{figure}

To simulate Faraday rotation, we incorporated $n_e(r,\xi, z)$, $B_\varphi(r,\xi,z)$ and $\bm{\beta}_f(r,\xi,z)$ into a dielectric function
\begin{equation}
\epsilon_\pm/\epsilon_0=1-\frac{\omega_p^2}{\omega_{pr}(\omega_{pr}\pm\Omega_B)}
\label{dielectric}
\end{equation}
at probe frequency $\omega_{pr}$ for the probe's right$^{(+)}$- and left$^{(-)}$-circularly-polarized components.  Here $\omega_p=\sqrt{n_e e^2/\epsilon_0 m_e}$ is the local plasma frequency, $\Omega_B=eB_{-x}/m_e$ the local cyclotron frequency, and $B_{-x}$ the component of $\mathbf{B}_\varphi$ along $\mathbf{\hat{k}}_{pr} = -\mathbf{\hat{x}}$.  Relativistic flow of electrons in the bubble walls necessitates a relativistic correction to $\epsilon_\pm/\epsilon_0$.\cite{Bro09} In the side walls (position \textbf{A} in Fig.\,\ref{bubble_flow}), $\bm{v}_{f,A}$ opposes $\bm{v}_b$, so here electrons flow at $v_b - v_{f,A} \sim 0$ in the lab frame, where the measurement is made.  Near the back of the bubble (position \textbf{B} in Fig.\,\ref{bubble_flow}), on the other hand, $\bm{v}_{f,B}$ has a strong component perpendicular to $\bm{v}_b$, resulting in net lab frame velocity $v_{net}$ as high as $0.97c$ ($\gamma_f \sim 4$) on the bubble axis, from the vector sum of $\bm{v}_b$ and $\bm{v}_{f,b}$. Thus relativistically corrected density $\tilde{n}_e=n_e/\gamma_f$, probe frequency $\tilde{\omega}_{pr}=\gamma_f(1-\bm{\beta}\cdot\mathbf{\hat{k}})\omega_{pr}$ and magnetic field $\tilde{\mathbf{B}}_\varphi=(\mathbf{1}+\gamma_f^2\bm{\beta\beta})\cdot\mathbf{B}_\varphi/\gamma_f$ must be used in Eq.\,\ref{dielectric}.\cite{Bro09} This suppresses contributions to $\phi_{rot}$ from the bubble's back wall relative to those from side walls by a factor $\sim 1/\gamma_f^4$.  We then find rotation angle by integrating
\begin{equation}
d\phi_{rot}=\frac{\omega}{c}\left(\sqrt{\epsilon_-/\epsilon_0} -\sqrt{\epsilon_+/\epsilon_0}\right) ds
\label{method2}
\end{equation}
over probe path elements $ds$ through the bubble, which leads to Eq.\,(\ref{rotation_rest}) for $\Omega_B, \omega_p \ll \omega_{pr}$. 
%\begin{equation}
%d\tilde{\phi}_3=\frac{\tilde{\omega}_p^2\cdot\tilde{\Omega}_B}{2\tilde{\omega}^2c}dx,
%\label{method3}
%\end{equation}
%which ultimately leads to the expression of 

%%%%%%%%%%%%%%%%%%%%%%%%%%%%
%
%		FIG 5
%
%%%%%%%%%%%%%%%%%%%%%%%%%%%%

Because pump and bubble move orthogonally to each other, a given point in the probe propagates across the bubble profile at $45^\circ$ to its $x$ and $\xi$ axes.  Accordingly, the dielectric function of the bubble at fixed propagation distance $z$ is $\epsilon_r^\pm(x,y,\xi)=\sum_{i} \epsilon_r^\pm(x=x _i +\xi, y,\xi)$ along propagation path $x=\xi$.  Here, $\{x_i\}$ are pixels on the $x$ axis from the PIC simulation, and we neglect bubble evolution during the probe transit. We performed a 2D+1 finite-difference time-domain (FDTD) simulation to advance each probe point stepwise along this path, and to accumulate polarization rotation according to Eq.\,(\ref{method2}).

%Similar to the 2D simulation, the probe beam was linearly polarized in y-axis. The difference here is that, instead of using a thin plane wave, we added a pulse length to the probe beam where $\tau_{pr}=100 $ $fs$. 

Fig.\,\ref{2D+1_FDTD}(b) shows the result of the simulation for a probe that propagated through the bubble profile in Fig.\,\ref{2D+1_FDTD}(a). This figure represents the probe's polarization profile upon emerging from the bubble.  Straight, parallel  streaks of length $c\tau_{pr}$ (FWHM), each of width $\Delta y_{sim} = 0.15\,R_b \approx 3.4 \pm 0.4\,\mu$m (FWHM), centered at $y = \pm 0.8\,R_b$, and corresponding to peak Faraday rotations $\phi_{rot}^{(max)} = \mp 0.4^\circ$, are observed.   The $y$-location of the streaks coincides closely with the $y$-location ($y\approx \pm0.75R_b$) at which the relativistically flattened magnetic field $B_\varphi$ of the accelerating $e$-bunch at $\Delta\xi/R_b \approx 0.5$ intersects a dense ($n_e \approx 7n_0$) section of the bubble's side wall [see dashed green lines in Fig.\,\ref{2D+1_FDTD}(a)]. This suggests that this section, pervaded by $B_\varphi$ of the most numerous and energetic accelerating electrons, contributes most of the observed Faraday rotation.  The simulated amplitude, length and separation of the streaks are consistent with corresponding observed characteristics shown in Fig.\,\ref{data_1}(c1).  Fig.\,\ref{2D+1_FDTD}(c) shows the simulated $\phi_{rot}$ profile when we artificially set $\gamma_f = 0$.  An additional contribution to $\phi_{rot}$ then appears close to $y = 0$.  This represents a contribution from the extreme rear bubble wall, magnetized by less numerous, low-energy electrons trailing the main \textit{e}-bunch. It is suppressed because of the net relativistic flow of the dense back-wall electrons. 

The simulated streak width in Fig.\,\ref{2D+1_FDTD}(b) is much narrower than the observed FWHM ($\Delta y_{exp} \approx 30\,\mu$m) shown in Figs.\,\ref{data_1}(c1).  When propagated through an imaging system with vertical resolution $\Delta y_{\rm res} = 12.5\,\mu$m [see Fig.\,\ref{setup}(d),(e)] to a detector, simulated streak width increases to $\Delta y_{\rm eff} \approx [(\Delta y_{\rm sim})^2 + (\Delta y_{\rm res})^2]^{1/2} \approx 13\,\mu$m, as shown by red-dashed vertical line-outs of the simulated streaks in Fig.\,\ref{data_1}(c1).  This is still significantly narrower than observed streaks.  We confirmed that object plane displacements within the depth of field of the imaging system, and variations in bubble wall thickness over a range $1 - 10\,\mu$m (conserving total particle number) negligibly affected the simulated width of the streak images. 

A possible source of this discrepancy is that Texas PW pulses blow out plasma bubbles less completely than idealized Gaussian pulses assumed in the simulations,  leaving blurred bubble edges that widen Faraday rotation streaks.  This would occur as a result of self-focusing to a smaller $a_0$ than simulated pulses.  An observable signature of lower $a_0$, however, would be smaller bubble radius (see Eq.\,1), and smaller streak separation.  In fact, we observe the opposite.  Centroids of observed streaks in Fig.\,3 are consistently located on average at $y \approx \pm 23\,\mu$m.  Assuming this represents $0.8\,R_b$ as in the simulations, it implies $R_b \approx 29\,\mu$m, \textit{larger} than the range of $R_b$ seen in the probed $z$-range in the simulations.  

A more likely source of the discrepancy is that plasma electrons inject more continuously, and thus spread over a longer region of the bubble's axis, than in the simulations.  The prominent low-energy tail observed in all electron spectra [see Fig.\,3(c3)-(e3)], which contrasts with near dark-current-free quasi-monoenergetic peaks  seen in the simulations,\cite{Kalm10} supports this hypothesis.  The azimuthal field $B_\varphi$ of these electrons would permeate a longer longitudinal section of the bubble side wall.  Because this section slopes in the $yz$-plane, it would generate a wider Faraday rotation streak, provided $\phi_{rot}$ is not too severely suppressed by net relativistic flow of side wall electrons.  

To test this hypothesis, we repeated the simulations for an earlier stage of acceleration when trapped electrons were closer to the bubble's rear [see Fig.\,\ref{2D+1_FDTD}(d)], and their magnetic fields pervaded a more steeply-sloped section of the bubble's side wall [indicated by dashed green lines in Fig.\,\ref{2D+1_FDTD}(d)].  Although Fig.\,\ref{2D+1_FDTD}(d) represents the state of the bubble at a smaller $z$ than observed in the experiment, here we use it to model $\phi_{rot}$ at the observed $z$ induced by low-energy electrons injected later than those in the simulation.  The FDTD simulation then indeed yields a wider Faraday rotation streak, as shown in Fig.\,\ref{2D+1_FDTD}(e), even with electron flow in the bubble wall taken fully into account.  Fig.\,\ref{2D+1_FDTD}(f) shows the calculated $\phi_{rot}$ when this flow is neglected:  $\phi_{rot}$ occurs much closer to the $y=0$ axis, and much larger in magnitude, than observed [see Fig.\,\ref{data_1}(c)-(e)].  The observed streaks in Fig.\,\ref{data_1}(c1) thus appear to be superpositions of narrow, widely-separated streaks [Fig.\,\ref{2D+1_FDTD}(b)] induced by the quasimonoenergetic leading edge of the accelerating \textit{e}-bunch, and broad, more closely spaced streaks [Fig.\,\ref{2D+1_FDTD}(e)] induced by its low-energy tail. A qualitatively similar mechanism likely accounts for the equally wide [Fig.\,\ref{data_1}(d1)] or wider [Fig.\,\ref{data_1}(e1)] Faraday streaks observed for the other shots shown in Fig.\,\ref{data_1}. 
 
 \section{Conclusion}

We have demonstrated single-shot Faraday rotation diagnosis of GeV LPAs in $40\times$-lower-density plasma than previous experiments while still using a near-IR probe pulse, thus achieving higher feature resolution than, but similar rotation magnitude to, prior work.  Analysis of the separation and width of positive- and negative-rotation streaks helps to determine transverse bubble size, degree of blowout, and the longitudinal profile of accelerating electrons. This work can be extended by splitting the probe into several replicas and multiplexing the detection system to sample the bubble along its entire multi-cm propagation path in one shot.  This would open the possibility of observing evolution of both accelerator structure and accelerating \textit{e}-bunch.  Compression of the probe to few-fs duration could additionally enable longitudinal profiling of the accelerating \textit{e}-bunch.\cite{Buc11}  The simple, inexpensive techniques demonstrated here are applicable to particle-bunch-driven plasma wakefield accelerators in plasma of similar $n_e$,\cite{Lit14,Cor15} in which magnetic fields of both drive and witness would contribute to the signal, enabling more complete characterization of the bubble wall and evolving \textit{e}-bunches than was possible here. 

\section*{Supplementary Material}
See supplementary material for a description of the data analysis procedure for large Faraday rotations.  

\begin{acknowledgments}
This work was supported by U.S. Department of Energy (DOE) grants DE-SC0011617 and DE-SC0014043, and U.S. National Science Foundation grant PHY-2010435. K. W. acknowledges additional support from a DOE Computational Science Graduate Fellowship (DE-FG02-97ER25308).  We gratefully acknowledge support from the staff of the Texas Petawatt Laser, a member of the LaserNetUS Collaboration, and from Harriet Hardman for editing the manuscript.
\end{acknowledgments}

\section*{Data Availability Statement}
The data that support the findings of this study are available from the corresponding author upon reasonable request.

\end{document}